\documentclass[11pt,preprintnumbers,aps,amssymb,nofootinbib,amsmath]
{revtex4}
\usepackage{epsfig,epsf}
\usepackage{graphicx}
\usepackage{verbatim}
\usepackage{epstopdf}
\usepackage{bm} 
\newcommand{\beq}{\begin{equation}}
\newcommand{\beql}[1]{\begin{equation}\label{#1}}
\newcommand{\eeq}{\end{equation}}
\newcommand{\bea}{\begin{eqnarray}}
\newcommand{\eea}{\end{eqnarray}}

\def\eq#1{{(\ref{#1})}}
\def\fig#1{{Fig.~\ref{#1}}}

\def\sec#1{{Sec.~\ref{#1}}}

\newcommand{\as}{\alpha_s}
\newcommand{\bas}{\bar\alpha_s}

\def\b#1{\mathbf{#1}}

\begin{document}

\preprint{RBRC-731}

\title{Spectrum of diffractively produced gluons  in onium--nucleus collisions}

\author{Yang Li$\,^a$ and Kirill Tuchin$\,^{a,b}$\\}

\affiliation{
$^a\,$Department of Physics and Astronomy, Iowa State University, Ames, IA 50011\\
$^b\,$RIKEN BNL Research Center, Upton, NY 11973-5000\\}

\date{\today}


\begin{abstract}
We calculate spectrum of  diffractively produced  gluons in onium--heavy nucleus collisions at high energies. We show that it exhibits a characteristic dependence on nucleus atomic number $A$ and energy/rapidity. We argue that this dependence offers a unique possibility for determining the low-$x$ structure of nuclear matter. Applications to RHIC, LHC and EIC experimental programs are discussed. 
\end{abstract}

\maketitle

\section{Introduction}\label{sec:intr}

Diffractive dissociation is one of the most interesting processes in high energy QCD. It played a pivotal role in identifying early signatures of the gluon saturation  \cite{Gribov:1983tu,Mueller:1986wy,McLerran:1993ni,Jalilian-Marian:1997jx,Jalilian-Marian:1997gr,Jalilian-Marian:1998cb,Jalilian-Marian:1997dw,Kovner:2000pt,Iancu:2000hn,Iancu:2001ad,Iancu:2001md,Ferreiro:2001qy} in deep inelastic scattering (DIS) at HERA \cite{GolecBiernat:1998js,GolecBiernat:1999qd,Gotsman:1999vt,Gotsman:2000gb,Gotsman:2002zi,Levin:2002fj,Levin:2001pr,Kovchegov:1999kx,Bartels:2002cj}. While the measurements at HERA revealed the first indication that the gluon saturation has become an important effect, the possible measurements of diffractive dissociation in p(d)A collisions at RHIC and LHC as well as in DIS in the proposed  EIC collider will be able  to probe gluon densities deeply in the saturation region. This statement is supported by the recent phenomenological success of models based on gluon saturation that accurately describe the experimental data on total hadron multiplicities \cite{Kharzeev:2000ph,Kharzeev:2001gp,Kharzeev:2001yq,Kharzeev:2002ei}, inclusive gluon production \cite{Kovchegov:1998bi,Kovchegov:2001sc,Braun:2000bh,Dumitru:2001ux,Blaizot:2004wu,Kharzeev:2002pc,Kharzeev:2003wz,Kharzeev:2004yx,Baier:2003hr,Iancu:2004bx} and heavy quark production \cite{Kharzeev:2003sk,Gelis:2003vh,Tuchin:2004rb,Blaizot:2004wv,Kovchegov:2006qn,Tuchin:2007pf,Kharzeev:2005zr,Tuchin:2006hz}. This motivated us to calculate multiplicity of diffractively produced gluons in coherent diffraction of  onium on a heavy nucleus in a recent publication \cite{Li:2008bm}. We observed that the diffractive gluon multiplicity is very sensitive to the low-$x$ dynamics in onium.  On the other hand, it showed only a weak dependence on the gluon density in the nucleus. The reason is that the total cross section is dominated by soft gluon momenta which are not sensitive to the short-distance structure on the nuclear color field. Consequently, in the present paper we set to calculate the diffractive gluon spectrum. Diffractive gluon production in DIS has been discussed in many publications \cite{Wusthoff:1997fz,GolecBiernat:1999qd,Bartels:1999tn,Kopeliovich:1999am,Gotsman:1999vt,Kovchegov:2001ni,Munier:2003zb,Marquet:2004xa,GolecBiernat:2005fe,Marquet:2007nf, Kovner:2006ge,Kovner:2001vi}. The goal of our calculation is to calculate, for the first time,  the  diffractive gluon spectrum taking into account the low-$x$ gluon evolution in all rapidity intervals, i.e.\ in the rapidity interval between the onium and the emitted gluon and between the emitted gluon and the nucleus. We believe that this calculation opens a new avenue towards the phenomenological applications in pA and eA collisions. 

The paper is organized as follows. In \sec{sec:model} we briefly review the result for  the gluon spectrum given in our previous paper \cite{Li:2008bm}. Eq.~\eq{eq1} represents the cross section in terms of the  dipole density in onium $n_p(\b r, \b r',\b b, y)$ and the dipole--nucleus forward scattering amplitude $N(\b r, \b b , y)$. In \sec{sec:dipole-ev} we deliberate  about the behavior of $n_p(\b r, \b r',\b b, y)$ and 
$N(\b r, \b b , y)$ in various kinematic regions.  We turn to analysis of the gluon spectrum  for large dipoles $r>1/Q_s$ in \sec{big} and small ones $r<1/Q_s$ in \sec{small}. The results are summarized in \sec{sec:summary} where we also discuss possible phenomenological applications.

\section{Diffractive gluon production in onium--nucleus collisions}\label{sec:model}

\begin{figure}[ht]
      \includegraphics[height=8cm]{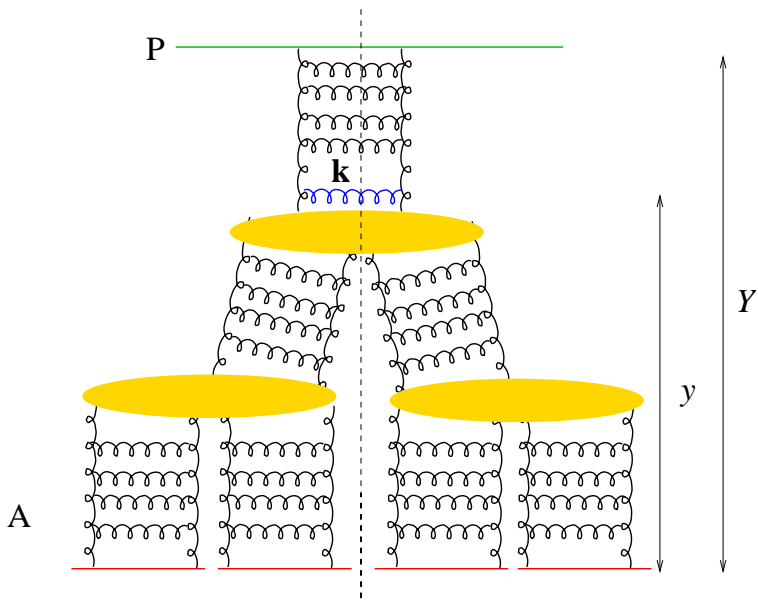}
\caption{Fan diagram for the diffractive gluon production in onium--nucleus collisions with maximal rapidity gap $Y_0=y$. The source of gluon multiplicity is the cut pomeron hanging out the onium P \cite{Li:2008bm}.}
\label{fig:diffract1}
\end{figure}
\begin{figure}[ht]
      \includegraphics[width=8cm]{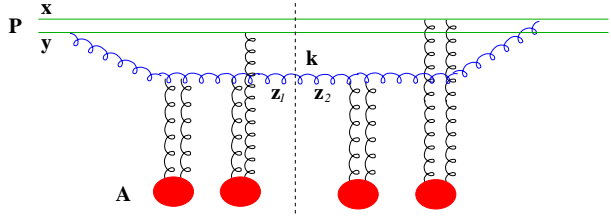}
\caption{One of the diagrams contributing to the diffractive gluon production at the quasi-classical level. Notations are detailed in text.}
\label{fig:not}
\end{figure}

Consider a process of diffractive gluon production in onium--nucleus scattering such that the rapidity gap equals the produced gluon rapidity $y$. The corresponding  fan diagram is displayed in \fig{fig:diffract1}.  In Ref.~\cite{Li:2008bm} we used the Mueller's dipole model  \cite{dip} to generalize the quasi-classical result of Kovchegov \cite{Kovchegov:2001sc} (derived independently in \cite{Kovner:2006ge}) by including the quantum evolution effects. 
The method is based on the principal idea of the dipole model that, due to the large difference between the coherence length of   the low-$x$ gluons in the onium light-cone ``wave-function" and the nuclear size, we can split  in the light--cone time the process of the low-$x$ evolution in onium and the instantaneous interaction. Indeed, the coherence length of the  low-$x$ gluons
is inversely proportional to $x$, whereas the size of the interaction region (in the nucleus rest frame) is $2R_A$. In the large $N_c$ approximation the onium wave function decomposes into a system of independent color dipoles. Up to  terms suppressed at  low $x$, dipole transverse size does not change in a course of interaction with the nucleus. We therefore, are able to write the cross section for the gluon production as a convolution of the onium dipole density and dipole  forward scattering amplitude. 
Let us introduce the following notations, see \fig{fig:not}: transverse coordinates of quark,  anti-quark, gluon in the amplitude and gluon in the c.c.\ amplitude are denoted by $\b x$, $\b y$, $\b z_1$, $\b z_2$ respectively; gluon transverse momentum is denoted by $\b k$.
In this notation the cross section takes the following form 
\bea\label{eq1}
\frac{d\sigma(k,y)}{d^2k\,
dy}&=&\frac{\as C_F}{\pi^2}\frac{1}{(2\pi)^2}\,\int d^2b\,d^2B\, \int
d^2r'\,  n_1(\b x-\b y,\b x'-\b y',\b B-\b b,Y-y)\, \nonumber\\
&&\times\bigg|\int d^2z_1\, e^{-i\b k\cdot \b z_1} \left(\frac{\b z_1-\b x'}{|\b z_1-\b x'|^2}- \frac{\b z_1-\b y'}{|\b z_1-\b y'|^2}\right)\,\left[ N(\b x'-\b y',\b b,y)\right.
\nonumber\\
&&\left. -N(\b x'-\b z_1,\b b,y)-N(\b
y'-\b z_1,\b b,y)+N(\b x'-\b z_1,\b b,y)N(\b y'-\b z_1,\b
b,y)\right]\bigg|^2\,,
\eea
where  $n_1(\b r,\b r',\b B-\b b,Y-y)$ is the dipole density and $N(\b r ',\b b,y)$ is the forward dipole--nucleus scattering amplitude. 
Function $n_1(\b x-\b y,\b x'-\b y',\b B-\b b,Y-y)$  has the meaning of the number of dipoles of size $\b x'-\b y'$ at rapidity $Y-y$ and impact parameter $\b b$ generated by evolution from the original dipole $\b x-\b y$ having rapidity $Y$ and impact parameter $\b B$ \cite{dip}. It satisfies the BFKL equation \cite{Kuraev:1977fs,Balitsky:1978ic} 
\bea\label{BFKL}
 &&\frac{\partial n_1(\b x-\b y,\b x'-\b y',\b b, y)}{\partial y}=\frac{\as N_c}{2\pi^2}\int d^2z \, \frac{(\b x-\b y)^2}{(\b x-\b z)^2(\b y-\b z)^2}\nonumber\\
&&
\big[
n_1(\b x-\b z,\b x'-\b y',\b b,y)+n_1(\b y-\b z,\b x'-\b y',\b b,y)-n_1(\b x-\b y,\b x'-\b y',\b b,y)\big]\,,
\eea
with the initial condition
\beq\label{in.cond}
n_1(\b r,\b r',\b b,0)=\delta(\b r-\b r')\,\delta(\b b)\,,
\eeq
where we denoted $\b r=\b x-\b y$ and $\b r'=\b x'-\b y'$.

The forward elastic dipole--nucleus scattering amplitude satisfies the nonlinear BK equation \cite{Balitsky:1995ub,Kovchegov:1999yj}
 \bea\label{BK}
 &&\frac{\partial N(\b x-\b y,\b b, y)}{\partial y}=\frac{\as N_c}{2\pi^2}\int d^2z \, \frac{(\b x-\b y)^2}{(\b x-\b z)^2(\b y-\b z)^2}
 \big[
N(\b x-\b z,\b b,y)\nonumber\\
&&
+N(\b y-\b z,\b b,y)-N(\b x-\b y,\b b,y)-N(\b x-\b z,\b b,y)N(\b y-\b z,\b b,y)
\big]\,,
\eea
with the initial condition given by  \cite{Mue}
\beq\label{NQ}
N(\b r,\b b,0)= 1-e^{ -\frac{1}{8}\b r^2\,
Q_{s0}^2}\,.
\eeq
The \emph{gluon} saturation scale is given by
\beq\label{Qsat}
Q_{s0}^2=\frac{4\pi^2\as N_c}{N_c^2-1}\,\rho\, T(\b b)\, xG(x,1/\b
r^2)\,,
\eeq
where $\rho$ is the nuclear density, $T(\b b)$ is the nuclear thickness function as a function of the impact parameter $\b b$. In the following we will assume for notational brevity that  the nuclear profile is cylindrical. An explicit impact parameter dependence, which are required in the numerical analysis, can be easily restored in the final expressions. Accordingly, it is convenient to proceed by defining the quantity
\beq\label{defnp}
n_p(\b r,\b r',y)=\int d^2b\,  n_p(\b r,\b r',\b b, y)
\eeq
which  satisfies the BFKL equation \eq{BFKL} with the initial condition
\beq\label{defnpic}
n_p(\b r,\b r',0)=\delta(\b r-\b r')\,.
\eeq

For the following calculations it is convenient to cast \eq{eq1} in a different form extracting the explicit dependence on $\b r'$. First, we change the integration variable $\b w=\b z_1-\b y'$. Then, introduce the following transverse vector
\bea\label{Igen}
\b I(\b r',\b k,y)&=&\int d^2w\, e^{-i\b k\cdot \b w} \left(\frac{\b w-\b r'}{|\b w-\b r'|^2}- \frac{\b w}{\b w^2}\right)\nonumber\\
&&\times\left[ N(\b  r',\b b,y)-N(\b w-\b r',\b b,y)-N(\b
w,\b b,y)+N(\b w-\b r',\b b,y)N(\b w,\b b,y)\right]\,.
\eea
Using \eq{Igen}, \eq{eq1} can be rendered as
\beq\label{main}
\frac{d\sigma(k,y)}{d^2k\,
dy}=\frac{\as C_F}{\pi^2}\frac{1}{(2\pi)^2}\,\int d^2b\,d^2B\, \int
d^2r'\,  n_1(\b r,\b r',\b B-\b b,Y-y)\,  |\b I(\b r',\b k,y)|^2\,,
\eeq
Now, contribution of the first term in the round brackets of \eq{Igen} can be written as 
\bea\label{Igen2}
&&\int d^2w\, e^{-i\b k\cdot w} \frac{\b w-\b r'}{|\b w-\b r'|^2}\left[ N(\b r',\b b,y)-N(\b w-\b r',\b b,y)-N(\b w,\b b,y)+N(\b w-\b r',\b b,y)N(\b w,\b b,y)\right]=\,\nonumber\\
&& -
\int d^2w\, e^{i\b k\cdot (\b w-\b r')} \frac{\b w}{\b w^2}\left[ N(\b r',\b b,y)-N(\b w-\b r',\b b,y)-N(\b w,\b b,y)+N(\b w-\b r',\b b,y)N(\b w,\b b,y)\right]\,
\eea
where we changed the integration variable  $\b w-\b r'\to -\b w$ and used the fact that the amplitude depends only on the dipole size (and not on direction). Defining a new scalar function $Q(\b r', \b k, y)$ as
\bea\label{Q}
&&Q(\b r', \b k, y)=\nonumber\\
&&
-\int d^2w\, e^{i\b k\cdot \b w} \frac{1}{w^2}\left[ N(\b r',\b b,y)-N(\b w-\b r',\b b,y)-N(\b w,\b b,y)+N(\b w-\b r',\b b,y)N(\b w,\b b,y)\right]\,.
\eea
and using \eq{Igen2} we write \eq{Igen} in the following form
\beq\label{iq}
\b I(\b r',\b k,y)=-e^{-i\b k\cdot \b r'}\, i \nabla_{\b k}Q(\b r',\b k,y)+i\nabla_{\b k} Q^*(\b r',\b k,y)\,.
\eeq
Consequently, 
\beq\label{i2}
|\b I(\b r',\b k,y)|^2=  2|\nabla_{\b k}Q(\b r',\b k,y)|^2-e^{-i\b k\cdot \b r'}(\nabla_{\b k}Q(\b r',\b k,y))^2-e^{i\b k\cdot \b r'}(\nabla_{\b k}Q^*(\b r',\b k,y))^2\,.
\eeq
In the region where   $Q(\b r', \b k, y)$ is a real function, we can render \eq{i2} as 
\beq\label{i2re}
|\b I(\b r',\b k,y)|^2= 4\,\sin^2\left(\frac{\b k\cdot\b r'}{2}\right)\,(\nabla_{\b k}Q(\b r', \b k, y))^2\,, \quad \mathrm{when} \,\,\, Q(\b r',\b k,y)\in \Re \,.
\eeq
In terms of $n_p(\b r, \b r', y)$, \eq{main} reads
\beq\label{main2}
\frac{d\sigma(k,y)}{d^2kdy} = \frac{\as C_F}{\pi^2}\frac{1}{(2\pi)^2}\,S_A\int d^2r' \, n_p(\b r, \b r', Y-y)\, |\b I(\b r',\b k,y)|^2\,,
\eeq
where $S_A$ is the cross sectional area of the interaction region. At $y=0$ this expression reduces to the quasi-classical formula derived in \cite{Kovchegov:2001ni,Kovner:2001vi,Kovner:2006ge}.

\begin{figure}[ht]
      \includegraphics[height=8cm]{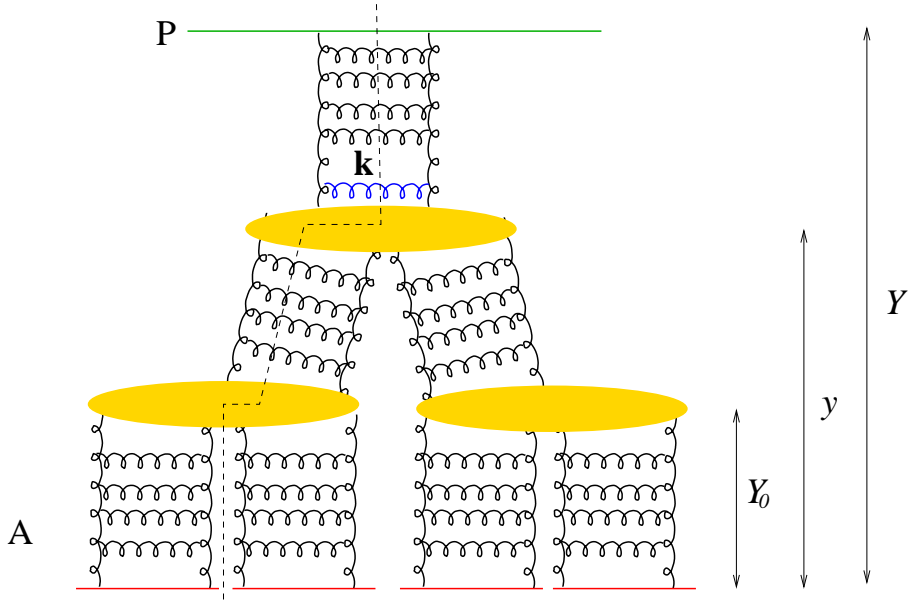}
\caption{Fan diagram for the diffractive gluon production in pA collisions with rapidity gap $Y_0$ smaller than the gluon rapidity $y$.}
\label{fig:diffract2}
\end{figure}

So far we have been concentrating on a case in which the rapidity of the produced gluon $y$ coincides with the rapidity gap $Y_0$ in a diffractive event. In this case the diffractive scattering amplitude $N_D(\b r, \b b, y, Y_0)$ coincides with the square of the forward elastic scattering amplitude $N(\b r, \b b, y)$. This case has the most phenomenological interest (since the invariant mass $M$ of the produced system is dominated by ``slow" gluons $M^2\approx k^2/x$). Still, at high enough luminosity a single hadron spectrum measurements should become possible. Therefore, a question may arise about the diffractive production of a gluon with $y> Y_0$. 
Such process is  shown in \fig{fig:diffract2}. 
In this case, the amplitude $N(\b r, \b b, y)$ must be replaced by the off-forward diffractive dipole amplitude which explicitly depends on a coordinate of  quark or anti-quark of parent dipole and the coordinates of the emitted gluon in the amplitude and in the c.c.\ one.\footnote{In the case of inclusive gluon production, an off-forward amplitude  was discussed in  \cite{JalilianMarian:2004da}.}  Investigation of properties of the off-forward diffractive amplitude would lead us astray of the main subject of this paper and hence will be discussed  elsewhere. Let us only note here that in those cases when the coordinate of the gluon is the same on both sides of the cut, the  off-forward diffractive amplitude reduces to the more familiar forward diffractive amplitude $N_D(\b r,\b b,y;Y_0)$. Since $N_D(\b r,\b b,y;Y_0)$ contains information about all possible pomeron cuts shown in \fig{fig:diffract2} it can serve as a \emph{phenomenological model} for the yet unknown off-forward diffractive amplitude. In this case 
the cross section for the diffractive gluon production takes form:
\beq\label{main3}
\frac{d\sigma^{pA}(k,y)}{d^2kdy} = \frac{\as C_F}{\pi^2}\frac{1}{(2\pi)^2}S_A\int d^2r' \, n_p(\b r', Y-y)\, |\b I(\b r',\b k,y;Y_0)|^2\,,
\eeq
where now in place of \eq{iq} and \eq{Q} we write
\beq\label{iq3}
\b I(\b r',\b k,y;Y_0)=-e^{-i\b k\cdot \b r'}\, i \nabla_{\b k}Q(\b r',\b k,y;Y_0)+i\nabla_{\b k} Q^*(\b r',\b k,y;Y_0)\,,
\eeq
and
\bea\label{Q3}
Q(\b r', \b k, y;Y_0)&=&-\int d^2w\, e^{i\b k\cdot \b w} \frac{1}{w^2}
\left[ N^\frac{1}{2}_D(\b r',\b b,y;Y_0)-N^\frac{1}{2}_D(\b w-\b r',\b b,y;Y_0)\right.\nonumber\\
&&
\left. \,-N^\frac{1}{2}_D(\b w,\b b,y;Y_0)+N^\frac{1}{2}_D(\b w-\b r',\b b,y;Y_0)N^\frac{1}{2}_D(\b w,\b b,y;Y_0)\right]\,.
\eea
Amplitude $N_D(\b r,\b b,y;Y_0)$ equals the cross section of single diffractive dissociation of a dipole of transverse size $\b r$, rapidity $y$ and impact parameter $\b b$ on a target nucleus. It satisfies the Kovchegov--Levin  evolution equation \cite{Kovchegov:1999ji}
\bea\label{BL}
&&\frac{\partial N_D(\b x-\b y,\b b, y;Y_0)}{\partial y}\nonumber\\
&&=\frac{2\as C_F}{\pi^2}\int d^2z \, \left[ \frac{(\b x-\b y)^2}{(\b x-\b z)^2(\b y-\b z)^2}-2\pi \delta(\b y-\b z)
 \ln(|\b x-\b y|\Lambda)\right] N_D(\b x-\b z,\b b,y;Y_0)
\nonumber\\
&+&
\frac{\as C_F}{\pi^2}\int d^2z \frac{(\b x-\b y)^2}{(\b x-\b z)^2(\b y-\b z)^2}
\left[N_D(\b x-\b z,\b b,y;Y_0)N_D(\b y-\b z,\b b,y;Y_0)\right.
\nonumber\\
&-& \left. 4 N_D(\b x-\b z,\b b,y;Y_0)N(\b y-\b z,\b b,y)+
2N(\b x-\b z,\b b,y)N(\b y-\b z,\b b,y)\right]\,,
\eea
with the initial condition
\beq\label{indif}
N_D(\b r,\b b,y=Y_0;Y_0)=N^2(\b r, \b b , Y_0)\,.
\eeq
Diffractive gluon production of the kind shown in \fig{fig:diffract2} requires a dedicated study while in this paper we concentrate on the case $y=Y_0$.

\section{Dipole evolution in onium and nucleus}\label{sec:dipole-ev}

\subsection{Dipole evolution in onium}

Dipole evolution in onium is encoded in the function $n_p(\b r, \b r', y)$
and is determined by solving  the BFKL equation \eq{BFKL} with the initial condition \eq{in.cond}. The result reads
\beq\label{np}
n_p(\b r,\b r',y)=\frac{1}{2\pi^2r'^2}\int_{-\infty}^\infty d\nu \, e^{2\bas\chi(\nu)y}\, (r/r')^{1+2i\nu}\,,
\eeq
where 
$\bas = \as N_c/\pi$ and 
\beq\label{chi}
\chi(\nu)=\psi(1)-\frac{1}{2}\psi(\frac{1}{2}-i\nu)-\frac{1}{2}\psi(\frac{1}{2}+i\nu)\,,
\eeq
with $\psi(\nu)$ being the digamma function
\beq\label{digamma}
\psi(\nu)=\frac{\Gamma'(\nu)}{\Gamma(\nu)}\,.
\eeq

There are several cases when the integral \eq{np} can be done analytically. 
Expansion near the maximum of $\chi(\nu)$ corresponds to the leading-logarithmic approximation. In this case we have 
\beq\label{npLLA}
n_p(\b r,\b r',y)_{LLA}\approx \frac{1}{2\pi^2rr'} \sqrt{\frac{\pi}{14\zeta(3)\bas y}}
e^{(\alpha_P-1)y}\, e^{-\frac{\ln^2(r'/r)}{14\zeta(3)\bas y}}\,,\quad \as y\gg \ln^2(r/r')\,,
\eeq
where $\alpha_P-1=4\bas \ln 2$.
Alternatively, we can expand $\chi(\nu)$ near one of its two symmetric poles   at $2i\nu =\pm 1$. This corresponds to the double logarithmic approximation depending on the relation between $r$ and $r'$. The results for the dipole density  read as follows 
\beq\label{npDLA}
n_p(\b r,\b r',y)_{DLA}\approx \frac{r^2}{4\pi^{3/2}r'^4}\frac{(2\bas y)^{1/4}}{\ln^{3/4}(r'/r)}\, e^{2\sqrt{2\bas y\ln(r'/r)}}\,,\quad r<r'\,,\quad \ln(r'/r)\gg \as y\,.
\eeq
and 
\beq\label{npDLA2}
n_p(\b r,\b r',y)_{DLA}\approx \frac{1}{4\pi^{3/2}r'^2}\frac{(2\bas y)^{1/4}}{\ln^{3/4}(r/r')}\, e^{2\sqrt{2\bas y\ln(r/r')}}\,,\quad r>r'\,,\quad \ln(r/r')\gg \as y\,.
\eeq

\subsection{Dipole evolution in a heavy nucleus}

In the region $rQ_s\ll 1$ where the forward elastic dipole--nucleus scattering amplitude $N(\b r,\b b, Y)$ satisfies the BFKL equation it can be calculated similarly to the dipole density of the previous subsection. The initial condition in this case is specified by \eq{NQ} expanded at small dipole sizes to the leading order. The result is
\beq\label{nag}
N(\b r,\b b, y)_{LT}=\frac{1}{8\pi}\int_{-\infty}^\infty d\nu\, e^{2\bas\chi(\nu)y}\, (rQ_{s0})^{1+2i\nu}\,\frac{1+(1-2i\nu)\ln\frac{Q_{s0}}{\Lambda}}{(1-2i\nu)^2}\,.
\eeq
Analogously to the derivation of \eq{npLLA} we obtain in the leading logarithmic approximation
\beq\label{naLLA}
N(\b r,\b b, y)_{LLA}=\frac{rQ_{s0}}{8\pi}\sqrt{\frac{\pi}{14\zeta(3)\bas y}}\ln\left(\frac{Q_{s0}}{\Lambda}\right) \, e^{(\alpha_P-1)y}\, e^{-\frac{\ln^2(rQ_{s0})}{14\zeta(3)\bas y}}\,,\quad \as y\gg \ln^2\left(\frac{1}{rQ_{s0}}\right)\,,
\eeq
and in  the double logarithmic approximation  
\bea\label{naDLA}
N(\b r,\b b, y)_{DLA}=\frac{\sqrt{\pi}}{16\pi}\frac{\ln^{1/4}\left(\frac{1}{rQ_{s0}}\right)}{(2\bas y)^{3/4}}\,r^2Q_{s0}^2\,\left( 1+  \sqrt{\frac{2\as y}{\ln\frac{1}{rQ_{s0}}}} \, \ln\frac{Q_{s0}}{\Lambda}  \right) e^{2\sqrt{2\bas y\ln\frac{1}{rQ_{s0}}}}\,,&&\nonumber\\
 r<1/Q_{s0}\,,\quad \ln\frac{1}{rQ_{s0}}\gg \as y\,.\qquad &&
\eea

Behavior of the scattering amplitude deeply in the saturation region $rQ_s\gg 1$ 
can be found by noting, that with the logarithmic accuracy, the parent dipole $\b r$ tends to split into  two daughter dipoles $\b w$ and $\b r-\b w$ of different sizes: either $w\ll r \approx |\b w-\b r|$ or, symmetrically,  $|\b w-\b r|\ll r \approx w$. Both give equal contribution to the integral over $\b w$. Restricting ourself to the case $w\ll r$ and doubling the integral we write the BK equation as follows:
\beq\label{xyz}
\frac{\partial N(\b r,\b b, y)}{\partial y}\approx \frac{\as C_F}{\pi}\,2\,\int_{1/Q_s^2}^{r^2}\frac{dw^2}{w^2}\,[N(\b w,\b b, y)-N(\b w,\b b, y)N(\b r,\b b, y)]\,.
\eeq
Now, for the reason that in the saturation region, the amplitude $N(\b r,\b b, y)$ is close to unity we render  \eq{xyz} as 
\beq\label{bksat}
-\frac{\partial\{1- N(\b r,\b b, y)\}}{\partial y}\approx\frac{\as C_F}{\pi}\,2\,\int_{1/Q_s^2}^{r^2}\frac{dw^2}{w^2} \{ 1-N(\b r,\b b, y)\}=
\frac{2\,\as C_F}{\pi}\ln(r^2Q_s^2)\, \{1- N(\b r,\b b, y)\} \,.
\eeq
The saturation scale $Q_s(y)$ can be found by equating the argument of the exponent in \eq{naDLA} to a constant which yields \cite{Levin:1999mw,Bartels:1992ix}
\beq\label{b0}
Q_s(y)\approx Q_{s0}e^{2\bas y}\,.
\eeq
Introducing a new scaling variable $\tau= \ln (r^2Q_s^2)$ we solve \eq{bksat} and find the high energy limit of the forward scattering amplitude \eq{bksat} \cite{Bartels:1992ix,Levin:1999mw,Levin:2000mv,Levin:2001cv} (in the fixed coupling approximation). It reads
\beq\label{lt}
N(\b r, \b b, y)=1-S_0\, e^{-\tau^2/8}= 1-S_0\, e^{-\frac{1}{8}\ln^2(r^2Q_s^2)}\,,\quad r\gg \frac{1}{Q_s}\,.
\eeq
where  we approximated $C_F\approx N_c/2$ in  the large $N_c$ limit. $S_0$ is the integration constant. It determines the value of the amplitude at the critical line $r(y)=1/Q_s(y)$.


\section{Diffractive gluon spectrum: large onium $r>1/Q_s$}
\label{big}

Now the stage is set for calculation of the diffractive gluon spectrum in various kinematic regions. 
Let us first analyze the differential gluon production cross section in the case of scattering of large onium $r> 1/Q_s$. There are two interesting kinematic regions in this case depending on the relation between the gluon transverse momentum $k$ and the saturation scale $Q_s$. We consider these two cases separately. 

\subsection{Hard gluons $k>Q_s$}

To begin we need to calculate  the function $Q(\b r', \b k, y)$ given by \eq{Q}. Note, that in the region $w>1/k$ the integrand is a rapidly fluctuation function. Therefore, the dominant contribution to $Q(\b r',\b k, y)$ arises from dipole sizes $w<1/k$. Consider now three possible cases. (i) $r'<\frac{1}{k}<\frac{1}{Q_s}$. In this case 
splitting the integration region into two parts we write \eq{Q} as 
\beq\label{iva1}
Q(\b r',\b k,y)\approx \int_0^{r'}\frac{d^2w}{w^2}\,[1-N(\b r',\b b, y)]\, N(\b w, \b b , y)\,+\, \int_{r'}^{1/k}\frac{d^2w}{w^2}\, 2N(\b w, \b b , y)\,.
\eeq
In the second integral on the r.h.s. we  neglected $N(\b r', \b b,0)$  as compared to  $N(\b w, \b b,0)$ since the amplitude is an increasing function of the dipole size  and in most of the integration region  $w\gg r'$. To determine the kinematic region that gives the largest contribution we note that when $w\ll \frac{1}{Q_s}$ the amplitude scales as $N(\b w, \b b, y)\sim w^2$. It follows, that the first integral in the r.h.s.\ of \eq{iva1} is of order $r'^2Q_s^2$ whereas the second one is of order $Q_s^2/k^2$, i.e.\ the former is parametrically smaller than the latter. Thus,
\beq\label{iva2}
Q(\b r',\b k,y)\approx \int_{r'}^{1/k}\frac{d^2w}{w^2}\, 2N(\b w, \b b , y)\,,\quad r'<\frac{1}{k}<\frac{1}{Q_s}\,.
\eeq
We obtain for the gradient 
\beq\label{iva4}
\nabla_{\b k}Q(\b r', \b k, y)=-4\pi \frac{\hat{\b k}}{ k}\,N(k^{-1}\hat {\b k}, \b b , y)\,,\quad  r'<\frac{1}{k}<\frac{1}{Q_s}\,.
\eeq
Eq.~\eq{iva4} holds in the logarithmic approximation. Using \eq{i2re} we get
\beq\label{iva5}
|\b I( \b r',\b k, y)|^2=4\,\frac{(4\pi)^2}{k^2}\, N^2(k^{-1}\hat {\b k}, \b b , y)\,\sin^2\left( \frac{\b k\cdot \b r'}{2}\right)\,,\quad  r'<\frac{1}{k}<\frac{1}{Q_s}\,.
\eeq
 
In the  second case  (ii) $\frac{1}{k}<r'<\frac{1}{Q_s}$ and the third case (iii)  $\frac{1}{k}<\frac{1}{Q_s}<r'$ there is only one significant integration region yielding 
\beq\label{iva3}
Q(\b r',\b k,y)\approx [1-N(\b r',\b b, y)]\,\int_{0}^{1/k}\frac{d^2w}{w^2}\, N(\b w, \b b , y)\,,\quad \frac{1}{k}<r'<\frac{1}{Q_s}\,\,\mathrm{and}\,\,\frac{1}{k}<\frac{1}{Q_s}<r' \,.
\eeq
Therefore, 
\beq\label{iva4a}
\nabla_{\b k}Q(\b r', \b k, y)=-2\pi \frac{\hat{\b k}}{ k}\,N(k^{-1}\hat {\b k}, \b b , y)\, [1-N(\b r',\b b, y)]\,\,,\quad  \frac{1}{k}<r'<\frac{1}{Q_s}\,\,\mathrm{and}\,\,\frac{1}{k}<\frac{1}{Q_s}<r' \,.
\eeq
Substitution into  \eq{i2re} yields
\bea\label{iva5a}
|\b I( \b r',\b k, y)|^2=4\,\frac{(2\pi)^2}{k^2}\, N^2(k^{-1}\hat {\b k}, \b b , y)\,[1-N(\b r',\b b, y)]^2\,\sin^2\left( \frac{\b k\cdot \b r'}{2}\right)\,, &&
\nonumber\\
\quad \frac{1}{k}<r'<\frac{1}{Q_s}\,\,\mathrm{and}\,\,\frac{1}{k}<\frac{1}{Q_s}<r' \,.&&
\eea
To calculate the differential cross section \eq{main2} we now need to integrate over all possible dipole sizes $r'$ and orientations. Integration over the dipole orientations produces 
\bea\label{xss1}
\frac{d\sigma}{d^2k\, dy}=\frac{\as C_F}{\pi^3}\,S_A\,\frac{(4\pi)^2}{k^2}\, N^2(k^{-1}\hat {\b k}, \b b , y)\, \left\{ \int_0^{1/k} dr' r' n_p(\b r, \b r', Y-y)\, (1-J_0(k\, r'))
\right.&&\nonumber\\
\left. +\frac{1}{4}\int_{1/k}^{1/Q_s} dr' r' n_p(\b r, \b r', Y-y)\, (1-J_0(k\, r')) \right\}
\,.
\eea
We restricted integration over $r'$ to the region $r'<1/Q_s$ since otherwise the integrand is strongly suppressed by $[1-N(\b r', \b b, y)]^2\to 0$, see \eq{iva5a} with \eq{lt} or \eq{NQ}.
To determine the largest contribution to the integral on the r.h.s.\ of \eq{xss1} we 
use the fact that the Bessel function $J_0(x)\sim x^{-1/2}$ at $x\gg 1$ and $J_0(x)\approx 1-x^2/4$ at $x\ll 1$ and write the expression in the curly brackets as
\beq\label{xss2}
\int_0^{1/k} dr' r' n_p(\b r, \b r', Y-y)\, \frac{1}{4}k^2 r'^2+\frac{1}{4}\int_{1/k}^{1/Q_s} dr' r' n_p(\b r, \b r', Y-y)\,.
\eeq
 It follows from   \eq{npDLA} and \eq{npDLA2} that  $n_p(\b r, \b r', y)\sim r^2/r'^4$ at $r<r'$ and  $n_p(\b r, \b r', y)\sim 1/r'^2$ at $r>r'$. On that account,  we determine that the first  integral in \eq{xss2} is of order unity, whereas the second one is logarithmically enhanced by $\ln(k/Q_s)\gg 1$. A more accurate estimate is gained by substitution of \eq{npDLA2} and explicit integration over $r'$.  Introducing a new integration variable  $\zeta=2\sqrt{2\bas (Y-y)\ln\frac{r}{r'}}$ we have
\beq\label{xss3}
\int_{1/k}^{1/Q_s} dr' r' n_p(\b r, \b r', Y-y)\approx \frac{\sqrt{2}}{4\pi^{3/2}}\,
\int_{\zeta_*}^{\zeta_0}\,\frac{d\zeta}{\sqrt{\zeta}}\,e^\zeta=
\frac{\sqrt{2\pi}}{4\pi^{3/2}}\,\left[\mathrm{erfi}\left(\sqrt{\zeta_0}\right)
-\mathrm{erfi}\left(\sqrt{\zeta_*}\right)\right]\,,
\eeq
where $\mathrm{erfi}(z)$ is the imaginary error function defined as 
\beq\label{err1}
\mathrm{erfi}(z)=-i\,\mathrm{erf}(iz)\,,
\eeq
$\zeta_0=2\sqrt{2\bas (Y-y)\ln(rk)}$ and $\zeta_*=2\sqrt{2\bas (Y-y)\ln(rQ_s)}$.  In compliance with the double logarithmic approximation we must replace   the imaginary error function by its asymptotic form at $\zeta_0\gg 1$ given by 
 \beq\label{err2}
 \mathrm{erfi}(z)\approx \frac{1}{\sqrt{\pi}\,z}\,e^{z^2}\,,\quad z\gg 1\,.
\eeq
Hence, keeping in mind that $\zeta_0\gg \zeta_*$ we get
\beq\label{xss10}
\int_{1/k}^{1/Q_s} dr' r' n_p(\b r, \b r', Y-y)
\approx \frac{1}{4\pi^{3/2}}\,\frac{1}{\left(2\bas (Y-y)\ln(rk)\right)^{1/4} }\,e^{2\sqrt{2\bas (Y-y)\ln(rk)}}\,.
\eeq
As expected, this integral is independent of $Q_s$ since the integrand is a steeply increasing function of $\frac{1}{r'}$. 
Finally, the cross section is procured by plugging \eq{xss10} into \eq{xss1} 
\beq\label{xssA}
\frac{d\sigma}{d^2k\, dy}=\frac{\as C_F}{\pi^{5/2}}\,\frac{1}{k^2}\, S_A\,N^2(k^{-1}\hat {\b k}, \b b , y)\,\frac{1}{\left(2\bas (Y-y)\ln(rk)\right)^{1/4} }\,e^{2\sqrt{2\bas (Y-y)\ln(rk)}}\,,
\quad r>\frac{1}{Q_s}>\frac{1}{k}\,.
\eeq

\subsection{Soft gluons $k<Q_s$}

We are now turning to analysis of soft gluon production by large onium. As in the case of hard gluons we wish to calculate $Q(\b r', \b k, y)$ in three different cases. First case corresponds to (i) $r'<\frac{1}{Q_s}<\frac{1}{k}$, i.e.\ size of dipole emitting the triggered gluon  is smaller than any other scale in the problem. We have
\beq\label{ivb1}
Q(\b r', \b k, y)\approx \int_0^{r'}\frac{d^2w}{w^2}[1-N(\b r',\b b, y)]\, N(\b w, \b b, y)
+\int_{r'}^{1/Q_s}\frac{d^2w}{w^2}\,2\,N(\b w, \b b, y)
+\int_{1/Q_s}^{1/k}\frac{d^2w}{w^2}\,,
\eeq
where we used the properties of the amplitude $N(\b w, \b b, y)$ as discussed after \eq{iva1}. The three  integrals on the r.h.s.\ of \eq{ivb1} is of order $r'^2Q_s^2\ll 1$, 1 and $\ln\frac{Q_s}{k}\gg 1$ respectively. Evidently, the third one is dominating. Thus, 
\beq\label{ivb2}
Q(\b r', \b k, y)\approx 2\pi\ln\frac{Q_s}{k}\,
\eeq
implying that
\beq\label{ivb3}
|\b I( \b r',\b k, y)|^2= \frac{4(2\pi)^2}{k^2}\,\sin^2\left( \frac{\b k\cdot \b r'}{2}\right)\,, \quad r'<\frac{1}{Q_s}<\frac{1}{k}\,.
\eeq

In the second case (ii) $\frac{1}{Q_s}<r'<\frac{1}{k}$ there are also three relevant regions of integration
 \bea\label{ivb7}
Q(\b r', \b k, y)&\approx& \int_0^{1/Q_s}\frac{d^2w}{w^2}[1-N(\b r',\b b, y)]\, N(\b w, \b b, y)+
\int_{1/Q_s}^{r'}\frac{d^2w}{w^2}[1-N(\b r',\b b, y)]\, N(\b w, \b b, y)\nonumber\\
&& + \int_{r'}^{1/k}\frac{d^2w}{w^2}[1-N(\b r',\b b, y)]\,N(\b w, \b b, y)\,.
\eea
In the second and the third integral  $N(\b w, \b b, y)\approx 1$. The third integral is enhanced by $\ln\frac{1}{r'k}$ and anyway it is the only integral that depends on $k$. Therefore, using \eq{lt}
\beq\label{ivb8}
\nabla_{\b k} Q(\b r', \b k, y)\approx -2\pi \frac{\hat {\b k}}{k}  \,[1-N(\b r',\b b, y)]=
-2\pi \frac{\hat {\b k}}{k}\, S_0\, e^{-\frac{1}{8}\ln^2 (Q_s^2r'^2)}
\,.
\eeq
Consequently, 
\beq\label{ivb10}
|\b I( \b r',\b k, y)|^2=\frac{4(2\pi)^2}{k^2}\,S_0^2\,e^{-\frac{1}{4}\ln^2(Q_s^2r'^2)} \,\sin^2\left( \frac{\b k\cdot \b r'}{2}\right)\,, \quad \frac{1}{Q_s}<r'<\frac{1}{k}\,.
\eeq

The third case corresponds to (iii) $\frac{1}{Q_s}<\frac{1}{k}<r'$. There are now two relevant regions 
\beq\label{ivb33}
Q(\b r', \b k, y)\approx \int_0^{1/Q_s}\frac{d^2w}{w^2}[1-N(\b r',\b b, y)]\, N(\b w, \b b, y)
+\int_{1/Q_s}^{1/k}\frac{d^2w}{w^2}\,[1-N(\b r',\b b, y)]\, N(\b w, \b b, y)\,.
\eeq
The second integral is enhanced by $\ln\frac{Q_s}{k}$ and, apart from the lower limit of integration, is the same as the third integral in \eq{ivb7}. 
Evidently, the $k$ dependence of  $Q(\b r', \b k, y)$  is the same as in the case (ii),  implying that \eq{ivb10} holds in the case (iii) as well
\beq\label{ivb10a}
|\b I( \b r',\b k, y)|^2=\frac{4(2\pi)^2}{k^2}\,S_0^2\,e^{-\frac{1}{4}\ln^2(Q_s^2r'^2)} \,\sin^2\left( \frac{\b k\cdot \b r'}{2}\right)\,, \quad \frac{1}{Q_s}<\frac{1}{k}<r'\,.
\eeq

Essentially, what \eq{ivb10} and \eq{ivb10a} tell us is that the region $r'>\frac{1}{Q_s}$ does not contribute to the cross section for diffractive production of soft gluon by large onium. Thus, the only contribution  to the cross section stems from $r'<\frac{1}{Q_s}$. There are now two possibilities depending on the size $r$ of the incident onium: (a) $r>\frac{1}{k}>\frac{1}{Q_s}$ and (b) $\frac{1}{k}>r>\frac{1}{Q_s}$. However, in both cases $\frac{1}{Q_s}$ is the smallest size implying that  the leading contribution to the cross section is the same in both cases. Expanding the argument of sinus in \eq{ivb3} and substituting to \eq{main2} we have  
\beql{ivbsec1}
\frac{d\sigma}{d^2k\,dy}=\frac{\as C_F}{\pi^2}\frac{1}{(2\pi)^2}\,S_A\,
\int_0^{1/Q_s}d^2r' n_p(\b r, \b r', Y- y)\, \frac{4(2\pi)^2}{k^2} \frac{1}{4}(\b k\cdot \b r')^2\,.
\eeq
The dipole density is given by  \eq{npDLA2}. Notice that since the largest contribution to the integral arises from dipoles of size $r'\sim \frac{1}{Q_s}$ (the integrand increase rapidly with $r'$) we can approximate $\ln\frac{r}{r'}\approx \ln(rQ_s)$, neglecting contribution of very small dipole sizes $r'$. Thus
\beql{ivbsec2}
\frac{d\sigma}{d^2k\,dy}=\frac{\as C_F}{8\pi^{5/2}}\,\frac{S_A}{Q_s^2}\,\frac{(2\bas (Y-y))^{1/4}}{\ln^{3/4}(rQ_s)}\,e^{2\sqrt{2\bas (Y-y)\ln(rQ_s)}}\,,\quad 
r,\frac{1}{k}>\frac{1}{Q_s}\,,
\eeq
which holds for any relation between $r$ and $1/k$.

If we now wish to calculate the total cross section for diffractive gluon production at given rapidity $y$ we have to integrate \eq{xssA} and \eq{ivbsec2} over $d^2k$. Clearly, the leading contribution stems from the integral over soft gluons given by \eq{ivbsec2}. We attain 
\beql{ivbsec3}
\frac{d\sigma}{dy}=\frac{\as C_F}{8\pi^{3/2}}\,S_A\,\frac{(2\bas (Y-y))^{1/4}}{\ln^{3/4}(rQ_s)}\,e^{2\sqrt{2\bas (Y-y)\ln(rQ_s)}}\,,\quad r>1/Q_s\,,
\eeq
in complete agreement with the result obtained in our previous paper \cite{Li:2008bm}.

\section{Diffractive gluon spectrum: Small onium $r<1/Q_s$}\label{small}

We now consider scattering of small onium on a heavy nucleus. We will again consider separately the two cases of hard and soft gluons. Calculation are facilitated a lot since we have already derived the function $|\b I( \b r',\b k, y)|^2$, which embodies information about the gluon emission and subsequent elastic scattering of the two intermediate dipoles $\b w$ and $\b r'-\b w$ off the nucleus.

\subsection{Hard gluons $k>Q_s$}

Using  \eq{iva5} we obtain
\beql{jj1}
\frac{d\sigma}{d^2k\, dy}=\frac{\as C_F}{\pi}\,\frac{1}{(2\pi)^2}\,S_A\, \frac{4\,(4\pi)^2}{k^2}\, N^2(k^{-1}\hat {\b k}, \b b , y)\,\int_0^\infty dr'r' n_p(\b r, \b r', Y-y)\,(1-J_0(kr'))\,,
\eeq
where we integrated over orientation of the dipole $\b r'$. To proceed we have to specify the relationship between the onium size $r$ and the inverse gluon transverse momentum $\frac{1}{k}$. Assume that (a) $r<\frac{1}{k}<\frac{1}{Q_s}$. Then, integral over $r'$ can be divided into the following four regions: (i) $0<r'<r$, (ii) $r<r'<\frac{1}{k}$, (iii) $\frac{1}{k}<r'<\frac{1}{Q_s}$ and (iv) $\frac{1}{Q_s}<r'$. To estimate the integral in each of this regions we use the same procedure as before (it is explained after \eq{xss2}). We find the following parametric dependence of the integral  in these four regions: (i) $k^2r^2$, (ii) $k^2r^2\ln \frac{1}{kr}$, (iii) $k^2r^2$ and (iv) $r^2Q_s^2$. Region (ii) gives the largest contribution. We have
\beql{jj2}
\frac{d\sigma}{d^2k\, dy}=\frac{\as C_F}{\pi}\,\frac{1}{(2\pi)^2}\,S_A\, \frac{4\,(4\pi)^2}{k^2}\, N^2(k^{-1}\hat {\b k}, \b b , y)\,\frac{k^2}{4}\,\int_0^\infty dr'r'^3\, n_p(\b r, \b r', Y-y) \,,
\eeq
Upon substitution of \eq{npDLA} and changing to a new integration variable 
$\tilde \zeta= 2\sqrt{2\bas (Y-y)\ln \frac{r'}{r}}$ we reduce the integral over $r'$ to the imaginary error function as in \eq{xss3}. Following the same steps as those that led us to \eq{xss10} we derive 
\beql{jj3}
\frac{d\sigma}{d^2k\, dy}=\frac{\as C_F}{\pi^{5/2}}\,S_A\, r^2\,N^2(k^{-1}\hat {\b k}, \b b , y)\,\frac{1}{\left(2\bas (Y-y)\ln\frac{1}{kr}\right)^{1/4} }\,e^{2\sqrt{2\bas (Y-y)\ln\frac{1}{kr}}}\,,
\quad r<\frac{1}{k}<\frac{1}{Q_s}\,.
\eeq

Consider region (b) $\frac{1}{k}<r<\frac{1}{Q_s}$. Repeating the same analysis as above we conclude that the dominant logarithmic contribution originates from the region $\frac{1}{k}<r'<r$. Accordingly, we use \eq{npDLA2} for the dipole density and neglect the Bessel function in \eq{jj1}. Doing the integral as explained in \eq{xss3}--\eq{xss10} we write
\beql{jj4}
 \frac{d\sigma}{d^2k\, dy}=\frac{\,\as C_F}{\pi^{5/2}}\,S_A\, \frac{1}{k^2}\,N^2(k^{-1}\hat {\b k}, \b b , y)\,\frac{1}{\left(2\bas (Y-y)\ln(rk)\right)^{1/4} }\,e^{2\sqrt{2\bas (Y-y)\ln(rk)}}\,,
\quad \frac{1}{k}<r<\frac{1}{Q_s}\,.
\eeq

\subsection{Soft gluons $k<Q_s$}

In the case $r<\frac{1}{Q_s}<\frac{1}{k}$ formulas for $|\b I( \b r',\b k, y)|^2$ are given by \eq{ivb3},\eq{ivb10},\eq{ivb10a}. As was already mentioned, only small dipoles $r'<1/Q_s$ contribute in the $r'$ integral. We thus have two regions of integration: (i) $0<r'<r$ and (ii) $r<r'<\frac{1}{Q_s}$. The integral over the former is of order $r^2$ whereas over the latter it is of order $r^2\ln \frac{1}{rQ_s}$. That being the case we derive
\bea
 \frac{d\sigma}{d^2k\, dy}&=&\frac{\as C_F}{\pi}\frac{1}{(2\pi)^2}\,S_A\, \int_r^{1/Q_s}dr' r' n_p(\b r, \b r', Y-y)\, \frac{4\,(2\pi)^2}{k^2}\, \frac{k^2\, r'^2}{4}\label{kk1}\\
 &=& 
  \frac{\as C_F}{4\pi^{5/2}}\,S_A\, r^2\,\frac{1}{\left(2\bas (Y-y)\ln\frac{1}{rQ_s}\right)^{1/4} }\,e^{2\sqrt{2\bas (Y-y)\ln\frac{1}{rQ_s}}}\,,
\quad r<\frac{1}{Q_s}<\frac{1}{k}\,.\label{kk2}
\eea
The total cross section is again dominated by soft gluons. Integrating \eq{kk1} over $d^2k$ such that $k<Q_s$ we find
\beql{ttt}
\frac{d\sigma}{dy}=\frac{\as C_F}{4\pi^{3/2}}\,S_A\, Q_s^2\,r^2\,\frac{1}{\left(2\bas (Y-y)\ln\frac{1}{rQ_s}\right)^{1/4} }\,e^{2\sqrt{2\bas (Y-y)\ln\frac{1}{rQ_s}}}\,,
\quad r<\frac{1}{Q_s}\,,
\eeq
again in agreement with our previous result \cite{Li:2008bm}.

\section{Summary} \label{sec:summary}

\begin{figure}[ht]
\begin{tabular}{lr}
      \includegraphics[height=5.cm]{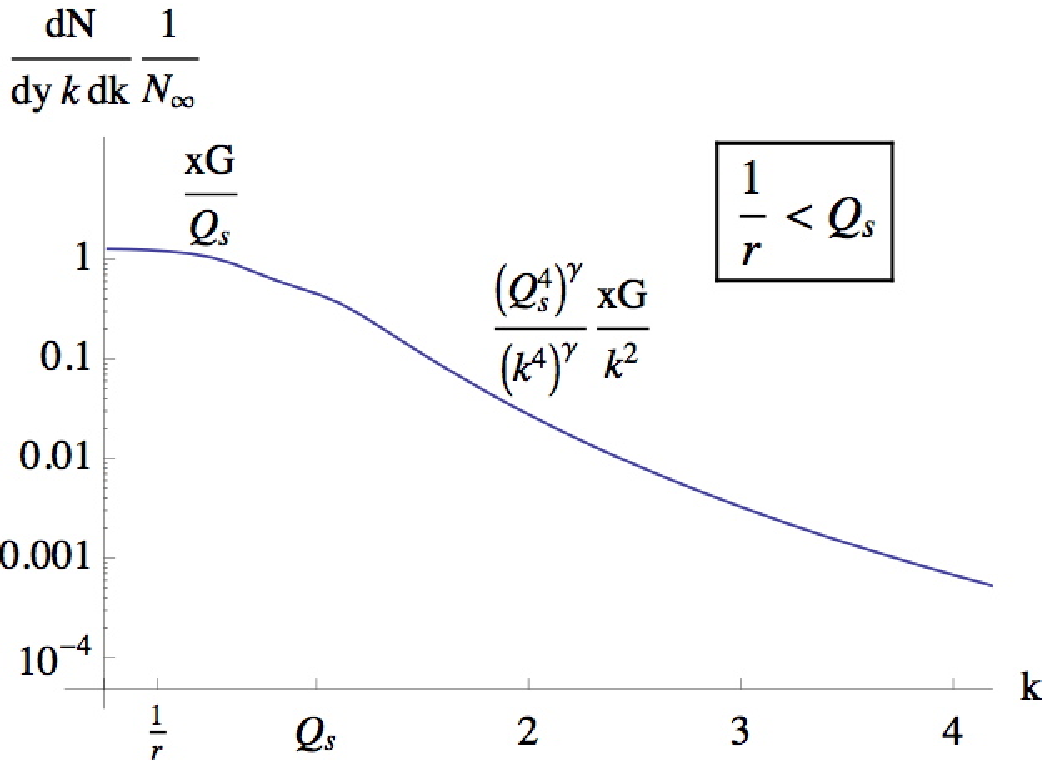}&
      \includegraphics[height=5.5cm]{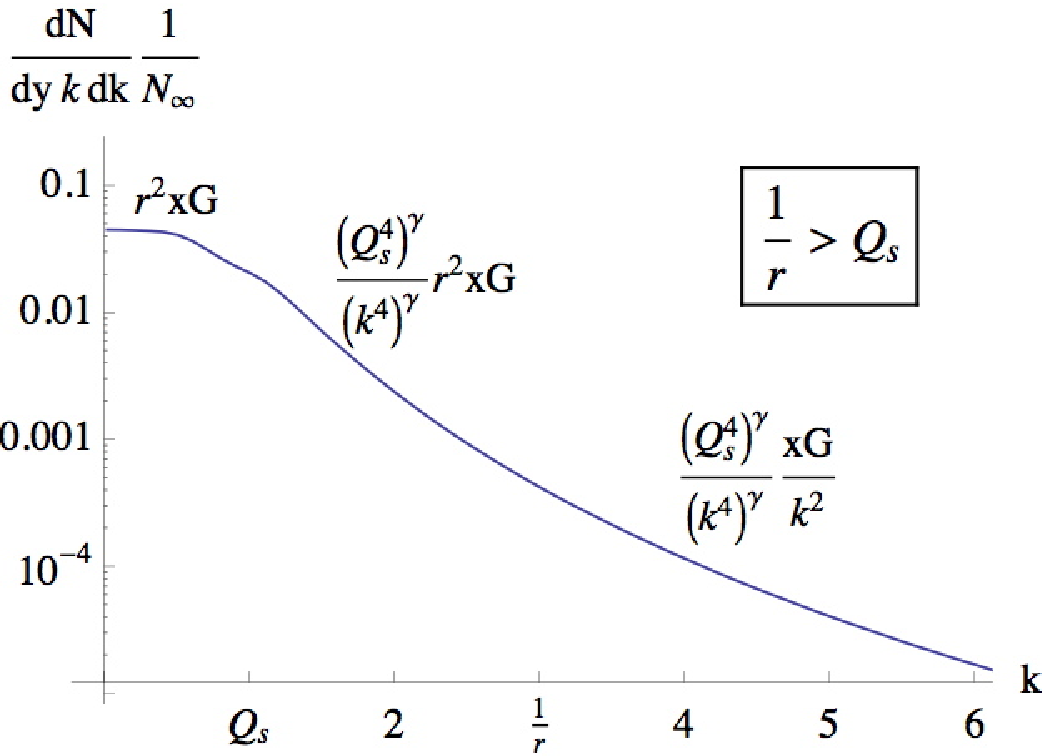}
 \end{tabular}     
\caption{Sketch of diffractive gluon spectrum as a function of transverse momentum $k$ in two cases $r>Q_s^{-1}$ and $r<Q_s^{-1}$. Qualitative behavior in both regions is also indicated. $xG(x=e^{y-Y})$ is a ``gluon distribution function" in onium, $\gamma$ is the anomalous dimension of the nuclear gluon distribution function and  $N_\infty$ is a normalization constant. }
\label{fig:rqs}
\end{figure}

The differential cross section for diffractive gluon production is given by  formulas \eq{xssA}, \eq{ivbsec2}, \eq{jj3}, \eq{jj4} and \eq{kk2}. We can see that there are five distinct kinematic regions, which are really six. The behavior of gluon spectrum in these regions is sketched  in \fig{fig:rqs}.
To make the figure self-contained we indicated an approximate  transverse momentum $k$ and the saturation scale $Q_s$  dependence.  $\gamma$ denotes the anomalous dimension of the nuclear gluon distribution. It varies from about unity at $k\gg Q_s^2/Q_{s0}$ to $\gamma\approx 1/2$ at $Q_s\lesssim k < Q_s^2/Q_{s0}$.
 \fig{fig:rqs} teaches us that by varying the incident onium size with respect to the saturation scale we obtain different behavior of the gluon spectrum as a function of transverse momentum. In DIS the typical onium size can be varied by means of triggering on the events with different photon virtuality. Depending on the relation between $k$, $Q_s$ and $r$ the gluon spectrum exhibits different pattern that allows a more direct measurement of the saturation scale $Q_s(y)$ (and hence the nuclear gluon density) than it is possible nowadays. The $k$-dependence of hadron spectra is of course significantly modified by the  
fragmentation process. On the other hand, dependence of hadron spectra on atomic number $A$ is the same as for the gluon spectrum (it arises from the $A$-dependence of $Q_s$, see \eq{Qsat}, \eq{b0}). The reason is that, as we explained in Introduction, the coherence length for the gluon production is much larger than the nucleus size, implying that the fragmentation process is independent of $A$. Consequently, $A$-dependence is a powerful tool in studying the nuclear gluon distribution.  Likewise, energy/rapidity dependence is independent of  details of fragmentation (see however \cite{Li:2007zzc}) and has been successfully used along with $A$-dependence for analysis of inclusive hadron production at RHIC. 
Therefore, energy/rapidity and atomic number dependence at different values of hadron transverse momenta  allows access to  information about the anomalous dimension $\gamma$, which is of crucial importance for understanding the transition region between the region of gluon saturation and the hard perturbative QCD. 

Similar arguments apply to the diffractive gluon production in pA collisions. In this case, however, there is a substantial uncertainty regarding the structure of the proton wave function. Diffractive gluon production in the case when the distance between the three pairs of valence quarks is about the same is strongly suppressed as compared to the case when the distance between one pair of quarks is much smaller than the distance between the other two pairs (quark - diquark configuration), see \cite{Li:2008bm}. In either case the $A$ and energy dependence are given by \fig{fig:rqs} (right or left panel). Since calculation of diffractive gluon production in pA collisions requires a substantial modeling of the proton wave function we intend to address it in a separate publication. 

To summarize, we calculated the spectrum of diffractively produced gluons at low-$x$ in onium--heavy nucleus collisions. In the forthcoming publications we are going to apply our results for calculation of the diffractive gluon production in DIS and pA collisions.

\acknowledgments
We would like to thank Yuri Kovchegov, Genya Levin and Jianwei Qiu for many informative discussions.  
The work of K.T. was supported in part by the U.S. Department of Energy under Grant No.\ DE-FG02-87ER40371. He would like to
thank RIKEN, BNL, and the U.S. Department of Energy (Contract No.\ DE-AC02-98CH10886) for providing facilities essential
for the completion of this work.


\end{document}